\documentstyle[12pt,a4]{article}
\makeatletter
\def\fmslash{\@ifnextchar[{\fmsl@sh}{\fmsl@sh[0mu]}}
\def\fmsl@sh[#1]#2{%
  \mathchoice
    {\@fmsl@sh\displaystyle{#1}{#2}}%
    {\@fmsl@sh\textstyle{#1}{#2}}%
    {\@fmsl@sh\scriptstyle{#1}{#2}}%
    {\@fmsl@sh\scriptscriptstyle{#1}{#2}}}
\def\@fmsl@sh#1#2#3{\m@th\ooalign{$\hfil#1\mkern#2/\hfil$\crcr$#1#3$}}
\makeatother
\global\arraycolsep=2pt 
\input{epsf}
\begin{document}
\thispagestyle{empty}
\begin{titlepage}
\begin{flushright}
{\bf TTP 95--22} \\
     Technion-PH-95-3 \\
     May 1995  \\
     hep-ph/9505288
\end{flushright}
\vspace{1cm}

\begin{center}
{\Large\bf PARTON HADRON DUALITY IN\vspace*{4mm} \\
           NONLEPTONIC $B$ HADRON DECAYS }
\end{center}
\vspace{0.8cm}

\begin{center}
{\sc Boris Blok} \vspace*{2mm} \\
{\sl Department of Physics, \\
     Technion - Israel Institute of Technology, Haifa 32000, Israel}
\vspace*{5mm} \\
{\sc Thomas Mannel}  \vspace*{2mm} \\
{\sl Institut f\"{u}r Theoretische Teilchenphysik, \\
     Universit\"at Karlsruhe \\
     D -- 76128 Karlsruhe, Germany.}
\end{center}
\vfill
\begin{abstract}
\noindent
Assuming so called global duality we argue that it is very likely
that local duality needed to obtain results for the hadronic width
of heavy meson decays within the $1/m_Q$ expansion holds. Hence, if
the discrepancy between experiment and the theory concerning
charm counting, the semileptonic branching fraction and the lifetimes
of $b$ hadrons persist, it may be taken as a hint at some
qualitatively new effect
in (nonperturbative) QCD or even as a new physics.
\end{abstract}
\end{titlepage}
\newpage
\section{Introduction}
The $1/m_Q$ expansion allows us to perform calculations of inclusive
heavy hadron decay rates in a model independent and QCD based framework
\cite{Bigi}.
Unlike for exclusive decays we are even able to deal with purely hadronic
processes using this method based on operator product expansion and
Heavy Quark Effective Theory. This in turn allows us to calculate
lifetimes and branching fraction for $b$ hardrons within this QCD
based approach.

The leading order of the $1/m_Q$ expansion is generically simply the
parton model expression. The corresponding decay rates in the parton
model are by now known for more than twenty years, and two problems
with the rates calculated in this way have been noticed since then.
Firstly, in the pure parton model, the semileptonic branching fraction
comes out to be too large as compared to the measurements and secondly
in the parton model all $b$ hadron lifetimes are the same and equal the
$b$ quark lifetime.

The hope was to fix these problems by including perturbative
and nonperturbative corrections originating from terms of order
$\alpha_s (m_Q)$ and $1/m_Q^2, 1/m^3_Q$.  It turned out
soon that the nonperturbative contributions are way too small to explain
the semileptonic branching fraction \cite{Baffling}
 as well as the lifetime difference
between the $B$ mesons and the $\Lambda_b$ baryon
\cite{BStalk}. In fact the
nonperturbative contributions are so small that they can safely be
ignored given the present experimental situation.

The purely perturbative corrections to the semileptonic decays have been
calculated some time ago \cite{Cabibbo}
 and were also found to be too small to explain
the data. The perturbative corrections to the hadronic decays
have also been calculated in the late seventies for the case of
vanishing masses of the final state quarks; again they are not sufficient
to explain the semileptonic braching fraction (see e.g. refs.
\cite{Petrarca,Baffling} for the recent reviews).

This was the motivation for Bagan and coworkers \cite{Bagan}
 to perform the full
${\cal O} (\alpha_s)$ calculation including the mass effects of the
final state $c$ quarks in the channel $b \to c \bar{c} s$. In fact,
Bagan et al.\ find a substantial enhancement for this channel, which,
together with the more modest enhancement in $b\rightarrow \bar cud$ channel
due to the same mechanism
almost explains the semileptonic branching fraction. This channel
has also been discussed previously in this context in \cite{Falketal},
where it was argued that the operator product expansion could fail due
to the small energy release in this channel, since the $c$ quarks are
heavy, or that parton hadron duality could fail due to resonances that
are close to the point where the hadronic spectral function is replaced
by the partonic one.

However, one cannot blame the problem with the semileptonic branching
fraction completely to the $b \to c \bar{c} s$ channel, since this would
increase this channel by something like (30 - 40)\%. This on the other
hand contradicts charm counting in $B$ decays, since this would give
for the number $n_c$ of $c$ and $\bar{c}$ quarks created per $B$ decay
$n_c \sim 1.3$ which is about two standard deviations away from the
experimentally observed value of $n_c = 1.07 \pm 0.07$ \cite{Glasrap}.
The QCD radiative corrections calculated in \cite{Bagan} also enhance
this channel; the full ${\cal O} (\alpha_s)$ calculation also gives
$n_c \sim 1.3$, again in  disagreement to charm counting.

Another problem for the $1/m_Q$ expansion is the lifetime ratio of the
$\Lambda_b$ baryon and the $B$ meson. While the $1/m_Q$ expansion predicts
a value in excess of 0.9 for the ratio $\tau(\Lambda_b) / \tau(B^+)$,
recent measurements yield
$\tau(\Lambda_b) / \tau(B^+) = 0.7 \pm 0.1$ \cite{Lambdab} which is a
deviation at the level of two standard deviations.

These two problems, namely the semileptonic branching fraction in
combination with charm counting and the lifetime of the $\Lambda_b$
baryon, have attracted considerable attention over the last time, although they
are not yet of convincing significance. Neither the nonperturbative nor
the perturbative corrections seem to be big enough to explain the problems
\cite{Baffling}. In particular, as it was mentioned above, although
the recently completed calculation of the complete ${\cal O} (\alpha_s)$
correction \cite{Bagan} is close to explain the discrepancy in the
semileptonic branching fraction, it does it by the price of generating
disagreement with charm counting.

All the arguments used in the $1/m_Q$ expansion are based on parton-hadron
duality and it has been suggested that a breakdown of duality is the
explanation of these discrepancies. It has been pointed out that unlike in the
calculation of the semileptonic branching fraction, where {\it global}
duality is needed, in the calculation of the hadronic width of the $B$
meson {\it local} duality needs to be invoked: In the calculation
of the semileptonic width only a weighted average of the sepectral function
of hadronic operators is replaced by the corresponding average of the
parton model expression, while for the hadronic width one replaces the
hadronic spectral function at a fixed kinematic point (i.e.\ locally)
by the partonic expression.

In this short note we shall assume that global duality holds and investigate
to what extend the hadronic spectral function may deviate from the paron model
expressions. Our arguments are necessarily qualitative, since it involves
the nonperturbative sector of QCD.  Intuitively one expects that the
final state quarks my form resonance states which may lead to a hadronic
spectral
function different from the partonic one. We shall use simple resonance
models to study the possible effects.

\section{Parton Hadron Duality}
We shall start from the effective Hamiltonian
describing nonleptonic $B$ decays, neglecting $b \to u$
transitions and penguins
\begin{eqnarray} \label{heff}
H_{eff} = \frac{G_F}{\sqrt{2}} V_{cb} \sum_{U = u,c} && \left[ C_2
  (\bar{b} \gamma_\mu (1-\gamma_5) c)
  (\bar{U} \gamma_\mu (1-\gamma_5) D_U) \right. \\
&& \left. + C_1
  (\bar{b} \gamma_\mu (1-\gamma_5) D_U)
  (\bar{U} \gamma_\mu (1-\gamma_5) c)\right] \nonumber
\end{eqnarray}
where $D$ is the Cabbibo-rotated down type quark field
\begin{eqnarray}
D_u &=& V_{ud}^* d + V_{us}^* s \\
D_c &=& V_{cd}^* d + V_{cs}^* s \nonumber
\end{eqnarray}
and $C_1$ and $C_2$ are the Wilson coefficients in the leading log
approximation
\begin{eqnarray}
C_1 &=& \frac{1}{2} \left[
        \left(\frac{\alpha_s(M_W)}{\alpha_s(m_b)}\right)^{\gamma_+} -
        \left(\frac{\alpha_s(M_W)}{\alpha_s(m_b)}\right)^{\gamma_-} \right] \\
C_2 &=& \frac{1}{2} \left[
        \left(\frac{\alpha_s(M_W)}{\alpha_s(m_b)}\right)^{\gamma_+} +
        \left(\frac{\alpha_s(M_W)}{\alpha_s(m_b)}\right)^{\gamma_-} \right]
\nonumber
\end{eqnarray}
Here
$$\gamma_+=\frac{6}{33-2N_f}=-\frac{1}{2}\gamma_-.$$
Using this Hamiltonian we may define the forward matrix element
\begin{equation}
T(q^2,vq) = \int d^4x \, e^{-iqx} \langle B(v) |
            H_{eff} (x) H_{eff} (0) | B(v) \rangle
\end{equation}
where $v = P_B/m_B$ is the velocity of the $B$ meson.

We shall use the correlator $T$ to study parton hadron duality. The
analytical structure of $T$ for complex $vq$ and fixed $q^2$
is given by the intermediate states
$| X \rangle$ that may be excited by $H_{eff}$
\begin{equation}
T(q^2,vq) = \int d^4 x \, (2\pi)^4 \delta^4 (m_B v - q - P_X)
| \langle B(v) | H_{eff} | X  \rangle | ^2  .
\end{equation}

The support of the function $T$ may easily be calculated. For a fixed
$q^2$, $T$ is nonvanishing in the interval
\begin{equation}
vq < \frac{1}{2m_B} (q^2 + m_B^2 - \mu^2) \mbox{ and }
vq > \frac{1}{2m_B} (q^2 + 2 m_B^2 + \mu^2)
\end{equation}
where $\mu$ is the mass of the lightest hadronic state which may appear in the
sum. In our case this state is a $D\pi$ state where both the $D$ and the
$\pi$ are at rest, so $\mu = m_D + m_\pi$.

In the parton model these intermediate states are three quark states
at the tree level, and the function $T$ may be calculated to be
\begin{eqnarray}
T(q^2,vq) &=& \left(\frac{G_F^2 m_b^4}{96 \pi^3} \right)
\langle B(v) | \bar{b} b | B(v)  \rangle N_c
\left( C_1^2 + C_2^2 + \frac{2C_1 C_2}{N_c} \right) V_{cb} \\
\nonumber
&& \left[ V_{ud}^* F_2 (m_c^2/Q^2) + V_{cs}^* F_3 (m_c^2/Q^2) \right]
\end{eqnarray}
where $Q = m_B v - q$. Furthermore,
we have neglected CKM suppressed terms proportional to $V_{us}$ and
$V_{cd}$ and
\begin{eqnarray}
F_2 (x) &=&(1-x^2)(1-8x+x^2)-12x^2 \ln (x)\Theta(1-x) \\
F_3 (x) &=& v(1-14x-2x^2-12x^3)
+24x^2(1-x^2)\ln\frac{1+v}{1-v}\Theta (1-4x),\\
&&  v=\sqrt{1-4x}
\end{eqnarray}
Note that the functions $F_2$ and $F_3$ have support in the intervals
\begin{equation}
vq < \frac{1}{2m_B} (q^2 + m_B^2 - m_c^2) .
\end{equation}
and
\begin{equation}
vq<\frac{1}{2m_B}(q^2+m_B^2-4m_c^2)
\end{equation}
respectively.

Parton hadron duality means that the correlator $T_{parton}$ calculated
in the parton model is ``on the average'' equal to the true hadronic
correlator $T_{hadron}$
\begin{equation} \label{ave}
\int d(vq) \, T_{parton} (q^2, vq) w(vq) =
\int d(vq) \, T_{hadron} (q^2, vq) w(vq), \quad q^2 \mbox{ fixed}
\end{equation}
where $w$ is some smooth weighting function having support over
some averaging interval. Close to the regions where
only the $D\pi$ or $J/\Psi \, K$ states
contribute the partonic and the hadronic results
have in general a quite different shape, and the interval over
which one needs to average in (\ref{ave}) is larger than if we consider
the two functions at higher invariant masses. THis situation has been
analyzed for the case of the total cross section $e^+ e^- \to $ hadrons
in \cite{PQW}.

For the calculation of
the total hadronic width we have to consider the correlator at the point where
the invariant mass of the final state hadrons is equal to the $B$ meson mass
\begin{equation}
\Gamma (B \to \mbox{hadrons}) \propto  T(0,0) ,
\end{equation}
and the common folklore  is that at this high invariant
mass one may savely shrink the interval of averaging in (\ref{ave}) to a point
and simply equate the partonic with the hadronic correlator. In other words,
we shall assume that ``global'' duality in the sense of a large
averaging interval holds and discuss under which circumstances we may shrink
the averaging interval to a single point, thereby going from ``global''
to ``local'' duality.

The situation in the nonleptonic decays has to be compared with the one
in semileptonic decays \cite{CDG,Bigi2}, where
the hadronic contribution is the product of the two hadronic currents.
In this case $q$ may be identified with the momentum transferred to the
leptons. While the analytical structure is quite similar (the lowest state
being a $D$ meson at rest) the integration over the neutrino momentum
introduces a smooth averaging such that parton hadron duality is guaranteed
in almost all phase space for the charged lepton, except for the endpoint,
where the energy of the charged lepton becomes maximal. Only if we would
consider
doubly differential distributions such that the invariant mass of the
final hadronic state would be fixed, a similar situation would occur as
in the nonleptonic decays, namely that the interval over which we average
shrinks to a point. In this sense one expects that the calculation of
the lepton spectrum as well as of the total semileptonic rate is more
reliable than the calculation of the total nonleptonic rate, since in the
semileptonic case only global duality is needed, while the nonleptonic
case rests on the assumption of global duality.

\section{Resonance Model}
Motivated by this reasoning we have considered the correlator for the
nonleptonic decays in the context of very simple model. The idea is to
identify possible contributions which could generate some structure of
the hadronic result which could result in a difference between
$T_{parton} (0,0)$ and $T_{hadron} (0,0)$ hence a violation of local
duality, although global duality holds.
As pointed out above, the
partonic tree level result is a smooth function of
$Q^2 = m_b^2 + q^2 - 2m_b(vq)$,
since it is obtained from a three particle phase space for the three
final state quarks. One possibility one may consider is that two of
these final state quarks form a resonance, corresponding e.g. to the
semiinclusive final processes $B \to D_{(s)}$+ anything or
$B \to J/\Psi$+ anything. In order to study the qualitative features of
this idea, we consider as a simple model for such a situation the
effective Hamiltonian
\begin{equation} \label{resmod}
H_{eff}^\prime = \frac{G_F}{\sqrt{2}} f_\phi
                 (\bar{b} \gamma_\mu (1-\gamma_5) q) \phi^\mu
\end{equation}
where $\phi_\mu$ is the field of a vector like resonance
of mass $M$ made of a quark-antiquark
pair, and $f_\phi$ is its coupling strength to the left handed
color singlet $b \to q$ current, including all the CKM matrix elements.
A similar model has been introduced by Stech and Palmer \cite{Stech}
to analyze semiinclusive nonleptonic $B$ decays.

It is a simple exercise to calculate the corresponding correlator.
The resonance contribution to $T$ near threshold  turns out to be
\begin{eqnarray} \label{rescon}
T(q^2, vq ) &=& \frac{G_F^2}{16 \pi} f_\phi^2
\langle B(v) | \bar{b} b | B(v) \rangle  \left(\frac{1}{Q^2}\right)^2 \\
\nonumber
&& \quad \sqrt{(Q^2 - (M - m_q)^2)(Q^2 - (M + m_q)^2)}
\Theta(Q^2 - (M - m_q)^2) \\ \nonumber
&& \quad \left[Q^2 - M^2 + m_q^2 + \frac{1}{M^2}
      ((Q^2 - m_q^2)^2 - M^4) \right]
\end{eqnarray}
Contributions of other resonances to the spectral function are smooth
since these resonances lie away from threshold, and we do not write
it explicitly.
Generically the phase space contributes the factor involving the square
root, and in the limit in which $M, Q^2
\gg m_q$ the r.h.s. of (\ref{rescon}) is
a smooth function even at threshold
\begin{eqnarray}
T(q^2, vq ) &=& \frac{G_F^2}{16 \pi} f_\phi^2
\langle B(v) | \bar{b} b | B(v) \rangle  \left(\frac{1}{Q^2}\right)^2 \\
\nonumber
&& \quad (Q^2 - M^2)^2 (Q^2 + M^2)
\Theta(Q^2 - M^2)
\end{eqnarray}

In our case $m_q$ is just the mass of the light strange quark.
This means that a scenario of a resonance as modelled in (\ref{resmod})
such that the thereshold is
close to the mass of the $B$ meson does not introduce any rapid variations
in the hadronic spectral function. Furthermore, since we assume global
duality in the sense of (\ref{ave}), one finds that the hadronic and
partonic spectral functions cannot be drastically different, at least
not in a scenario as moddelsed by the Hamiltonian (\ref{resmod}).
Hence we conclude that it is quite likely that the partonic
correlator $T$ is a good approximation to the hadronic,
even for the case of a resonance as we introduced above fairly
close to the mass of the $B$ meson.

Typically the correlators obtained from intermediate states with
two or more partons or hadrons are smooth functions of the variable
$Q^2 = m_B^2 + q^2 -2m_b (vq)$ and any resonance
formation involving at least a two particle intermediate state does not
introduce rapid variations of the correlator. For all these states we
may apply the same argument as above, leading to the statement that
it is quite likely that local duality holds. In this framwork, the only way
to introduce strong variations is if there is a relatively narrow
resonance $R_c$ leading to
a single particle intermediate state in the correlator $T$.

In order  not to get into conflict with charm counting, this resonance
state has to appear in the channel $b \to c \bar{u} d$, thereby enhancing
this channel by some amount, which then would explain the semileptonic
branching fraction without enhancing the charm created in $B$ decays.
In this case, the correlator would take the form
\begin{equation}
T(q^2,vq) = 2M\Gamma \frac{\langle B(v) | H_{eff} |R_c \rangle
\langle R_c | H_{eff} | B(v) \rangle}{(Q^2 - M^2)^2 + M^2\Gamma^2}
\end{equation}
where $M$ and $\Gamma$ are the mass and the width of $R_c$ respectively.
If this resonance would be sufficiently narrow and its mass sufficiently
close to the mass of the $B$ meson,  parton hadron duality
could still hold on the average, but the interval of averaging would simply
be too large to justify the calculation of the total hadronic width using
local duality.

However, this possibility is quite unlikely. First of all, it would not help
to explain the lifetime problem with the $\Lambda_b$ baryon, since such a
resonance state would increase the width of the $B$ mesons due to an increase
in the channel $b \to c \bar{u} d$,  and hence would make
their lifetime shorter compared to the parton model value. Assuming that there
is not such a resonance in the channel involving the baryons, this would
increase
the ratio $\tau(\Lambda_b) / \tau(B)$, which will make the discrepancy even
stronger.

Furthermore, the effect of such a resonance would be visible as well in the
semileptonic channel, if the resonance state would have isospin $I=1/2$.
However, one may avoid any influence on the semileptonic decays, if the
isospin of $R_c$ is $I = 3/2$.  This would
mean that in a quark model picture this state is a four quark state
consisting of a two quarks and two antiquarks.

However, the $b \to c\bar{u}d$ piece of the effective Hamiltonian (\ref{heff})
is an isospin triplet with $I_3 = -1$, and hence the hadronic decays of the
charged $B$ mesons will have only final states with isospin $I = 3/2$, while
the
final states of the nonleptonic decays of the neutral $B$ meson will be a
mixture of $I = 3/2$ and $I = 1/2$. Consequently an enhancement of the
$I = 3/2$ contribution through a resonance would make the lifetime of
the charged $B$ meson shorter than the one of the neutral states, which
is in conflict with the data on the lifetime ratio between charged and neutral
$B$ mesons, which indicate that the lifetimes are equal
$\tau (B^+) / \tau (B^0) = 1.009 \pm 0.069$ \cite{Glasrap}. Hence there is
not much room for such a resonance state and even if it would exist,
its contribution would be not sufficient to explain the semileptonic
branching fraction problem.

\section{Conclusions}
Even after more detailed calculations of the QCD radiative corrections
the situation concerning charm counting and the semileptonic branching
fraction remains inconclusive. A failure of local parton hadron duality
as it is needed to obtain results within the $1/m_Q$ expansion has been
considered as a possible solution of the problem, but we have argued
based on models that local parton hadron duality likely to hold, if
global duality is assumed.

Global duality, on the other hand, is a concept without which we could
not relate many theoretical results to data, which in turn are compatible
with the assumption of global duality. Hence the semileptonic branching
fraction problem and the charm counting remain open questions in view of
the present data. Either the data on charm counting
change, such that $n_c \sim 1.3$. In this case the problem is solved,
and such a result would also be in agreement with the recent calculation
of the full ${\cal O} (\alpha_s)$ corrections \cite{Bagan}. However,
if data remain as they are and the errors become smaller, than there have
to be either yet unknown effects in nonperturbative QCD, or the
discrepancies find an explanation in terms of physics beyond the
standard model \cite{kagan,shifman}. However, we find an explanation
in terms of QCD effects more likely, since physics beyond the standard
model involves scales much larger than the $B$ meson mass and hence
will affect the decays of the $b$ {\it quark} rather than the ones of the
hadrons, in other words, such an explanation cannot be the source of the
$\Lambda_b$ lifetime problem.

\section*{Acknowledgements}
The authors thank M. Gronau and M. Beneke and especially
 M. Shifman and  A. Vainshtein for very
usefull discussions. T.M. acknowledges the hospitality of the particle
theory group of the Technion, where this work has been initiated.
The research of B.B. was supported by the Israel Academy of Sciences
under the contract 938041 and by the Technion V.P.R. fund.

\end{document}